\documentclass[runningheads,fleqn]{svmult}
\usepackage{makeidx}   
\usepackage{graphicx}  
\usepackage{subeqnar}  
\usepackage{multicol}  
\usepackage{taphys}    
\makeindex             
\begin{document}
\title*{Interacting neural networks and cryptography}

%
\author{Wolfgang Kinzel\inst{1} \and Ido Kanter\inst{2}}
\authorrunning{Wolfgang Kinzel and Ido Kanter}
%
%
\institute{Institute for Theoretical Physics and Astrophysics,
  Universit\"at W\"urzburg, Am Hubland, 97074 W\"urzburg, Germany \and
  Minerva Center and Department of Physics, Bar-Ilan University, 52100
  Ramat-Gan, Israel}

\maketitle              

\begin{abstract}
  \index{abstract} Two neural networks which are trained on their
  mutual output bits are analysed using methods of statistical
  physics. The exact solution of the dynamics of the two weight
  vectors shows a novel phenomenon: The networks synchronize to a
  state with identical time dependent weights. Extending the models to
  multilayer networks with discrete weights, it is shown how
  synchronization by mutual learning can be applied to secret key
  exchange over a public channel.
\end{abstract}

\section{Introduction}
Neural networks learn from examples. This concept has extensively been
investigated using models and methods of statistical mechanics 
\cite{HeKrPa,EnVa}. A
''teacher'' network is presenting input/output pairs of high dimensional
data, and a ''student'' network is being trained on these data. Training
means, that synaptic weights adopt by simple rules to the input/output
pairs.

When the networks --- teacher as well as student --- have $N$ weights,
the training process needs of the order of $N$ examples to obtain
generalization abilities. This means, that after the training phase the
student has achieved some overlap to the teacher, their weight vectors
are correlated. As a consequence, the student can classify an input
pattern which does not belong to the training set. The average
classification error decreases with the number of training examples.

Training can be performed in two different modes: Batch and on-line
training. In the first case all examples are stored and used to minimize
the total training error. In the second case only one new example is
used per time step and then destroyed. Therefore on-line training may be
considered as a dynamic process: at each time step the teacher creates a
new example which the student uses to change its weights by a tiny
amount. In fact, for random input vectors and in the limit $N
\rightarrow \infty$, learning and generalization can be described by
ordinary differential equations for a few order parameters \cite{BiCa}.

On-line training is a dynamic process where the examples are generated
by a static network - the teacher. The student tries to move towards the
teacher. However, the student network itself can generate examples on
which it is trained. When the output bit is moved to the shifted input
sequence, the network generates a complex time series \cite{EiKaKeKi}. Such
networks are called bit (for binary) or sequence (for continuous
numbers) generators and have recently been studied in the context of
time series prediction \cite{KaKePrEi}.

This work on the dynamics of neural networks - learning from a static
teacher or generating time series by self interaction - has motivated us
to study the following problem: What happens if two neural networks
learn from each other? In the following section an analytic solution is
presented \cite{MeKiKa}, which shows a novel phenomenon: synchronization by
mutual learning. The biological consequences of this phenomenon are not
explored, yet, but we found an interesting application in cryptography:
secure generation of a secret key over a public channel.

In the field of cryptography, one is interested in methods to transmit
secret messages between two partners A and B. An opponent E who is able
to listen to the communication should not be able to recover the secret
message.

Before 1976, all cryptographic methods had to rely on secret keys for
encryption which were transmitted between A and B over a secret channel
not accessible to any opponent. Such a common secret key can be used,
for example, as a seed for a random bit generator by which the bit
sequence of the message is added (modulo 2).

In 1976, however,  Diffie and Hellmann found that a common secret key
could be created over a public channel accessible to any opponent. This
method is based on number theory: Given limited computer power, it is
not possible to calculate the discrete logarithm of sufficiently large
numbers \cite{St}.

Here we show how neural networks can produce a common secret key by
exchanging bits over a public channel and by learning from each other.

\section{Dynamic transition to synchronization}
\begin{figure}
\sidecaption
\includegraphics[width=.4\textwidth]{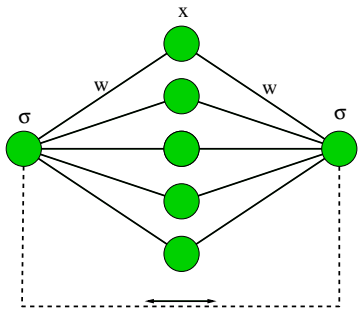}
\caption[]{Two perceptrons receive an identical input $\underline{x}$ and learn their mutual output bits $\sigma$.}
\label{per}
\end{figure}

Here we study mutual learning of neural networks for a simple model
system: Two perceptrons receive a common random input vector
$\underline{x}$ and change their weights $\underline {w}$ according to
their mutual bit $\sigma$, as sketched in Fig. \ref{per}. The output bit $\sigma$ of a single
perceptron is given by the equation

\begin{equation}\label{eins}
\sigma = \mbox{sign} (\underline{w} \cdot \underline{x})
\end{equation}
$\underline{x}$ is an $N$-dimensional input vector with components which
are drawn from a Gaussian with mean $0$ and variance $1$. $\underline{w}$
is a $N$-dimensional weight vector with continuous components which are
normalized,

\begin{equation}\label{zwei}
\underline{w} \cdot \underline{w}= 1
\end{equation}

The initial state is a random choice of the components $w_i^{A/B},
i=1, ... N$ for the two weight vectors $\underline{w}^{A}$ and
$\underline{w}^{B}$.  At each training step a common random input
vector is presented to the two networks which generate two output bits
$\sigma^{A}$ and $\sigma^B$ according to (\ref{eins}). Now the weight
vectors are updated by the perceptron learning rule \cite{BiCa}:

\begin{eqnarray}\label{drei}
\underline{w}^{A} (t+1) & = & \underline{w}^{A} (t) + \frac{\eta}{N} \underline{x} \sigma^B \; \Theta(-\sigma^A \sigma^B) \nonumber\\
\underline{w}^B (t+1) & = & \underline{w}^B (t) + \frac{\eta}{N} \underline{x} \sigma^{A} \; \Theta (-\sigma^{A} \sigma^B)
\end{eqnarray}
$\Theta(x)$ is the step function. Hence, only if the two perceptrons
disagree a training step is performed with a learning rate $\eta$. After
each step (\ref{drei}), the two weight vectors have to be  normalized.

In the limit $N \rightarrow \infty$, the overlap

\begin{equation}\label{vier}
R(t) = \underline{w}^{A} (t) \; \underline{w}^{B} (t)
\end{equation}
has been calculated analytically \cite{MeKiKa}.
The number of training steps $t$ is scaled as $\alpha = t/N$, and
$R(\alpha)$ follows the equation

\begin{equation}\label{fuenf}
\frac{d R}{d \alpha} = (R+1) \left( \sqrt{\frac{2}{\pi}} \; \eta(1-R) - \eta^2 \frac{\varphi}{\pi} \right)
\end{equation}
where $\varphi$ is the angle between the two weight vectors
$\underline{w}^{A}$ and $\underline{w}^B$, i.e. $R= \cos \varphi$. This
equation has fixed points $R=1, R=-1$, and

\begin{equation}\label{sechs}
\frac{\eta}{\sqrt{2 \pi}} = \frac{1- \cos \varphi}{\varphi}
\end{equation}

\begin{figure}
\sidecaption
\includegraphics[width=.6\textwidth]{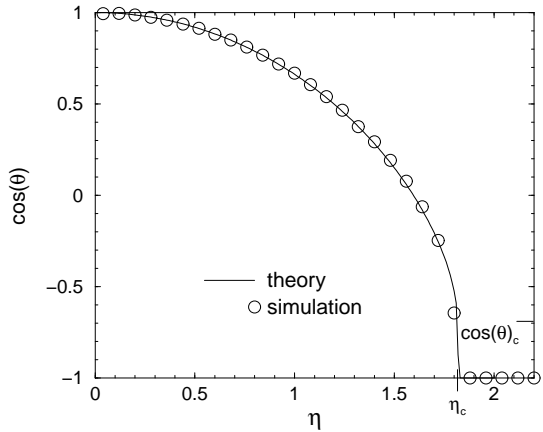}
\caption[]{
  Final overlap $R$ between two perceptrons as a function of learning
  rate $\eta$. Above a critical rate $\eta_c$ the time dependent
  networks are synchronized. From Ref. \cite{MeKiKa} }
\label{fig1}
\end{figure}
Fig. \ref{fig1} shows the attractive fixed point of \ref{fuenf} as a
function of the learning rate $\eta$. For small values of $\eta$ the two
networks relax to a state of a mutual agreement, $R \rightarrow 1$ for
$\eta \rightarrow 0$. With increasing learning rate $\eta$ the angle
between the two weight vectors increases up to $\varphi = 133^{\circ}$
for

\begin{equation}\label{sieben}
\eta \rightarrow \eta_c \cong 1.816
\end{equation}
Above the critical rate $\eta_c$ the networks relax to a state of
complete disagreement, $\varphi = 180^{\circ}, R= - 1$. The two weight
vectors are antiparallel to each other, $\underline{w}^{A} = -
\underline{w}^B$.

As a consequence, the analytic solution shows, well supported by
numerical simulations for $N = 100$, that two neural networks can
synchronize to each other by mutual learning. Both of the networks are
trained to the examples generated by their partner and finally obtain
an antiparallel alignment.  Even after synchronization the networks
keep moving, the motion is a kind of random walk on an N-dimensional
hypersphere producing a rather complex bit sequence of output bits
$\sigma ^{A} = - \sigma^B$ \cite{MeKiEiKa}.

\section{Random walk in weight space}
We want to apply synchronization of neural networks to cryptography. In
the previous section we have seen  that the weight vectors of two
perceptrons learning from each other can synchronize. The new idea is to use
the common weights $\underline{w}^{A} = - \underline{w}^B$ as a key for
encryption \cite{KaKiKa}. But two problems have to be solved yet: (i) Can an
external observer, recording the exchange of bits, calculate the final
$\underline{w}^{A} (t)$, (ii) does this phenomenon exist for discrete
weights? Point (i)  is essential for cryptography, it  will be discussed
in the following section. Point (ii) is important for practical
solutions since communication is usually based on bit sequences. 
It will be investigated in the following.

Synchronization occurs for normalized weights, unnormalized ones do not
synchronize \cite{MeKiKa}. Therefore, for discrete weights, we introduce a
restriction in the space of possible vectors and limit the components
$w_i^{A/B}$ to $2L+1$ different values,

\begin{equation}\label{acht}
w_i^{A/B}  \in \{ -L, -L +1, ... , L-1, L \}  
\end{equation}
In order to obtain synchronization to a parallel -- instead of an
antiparallel -- state $\underline{w}^{A} = \underline{w}^B$, we modify
the learning rule (\ref{drei}) to:

\begin{eqnarray}\label{neun}
\underline{w}^{A} (t+1) & = & \underline{w}^{A} (t) - \underline{x} \sigma^{A} \Theta (\sigma^{A} \sigma^B) \nonumber \\
\underline{w}^B (t+1) & = & \underline{w}^B (t) - \underline{x} \sigma^B \Theta (\sigma^{A} \sigma^B)
\end{eqnarray}
Now the components of the random input vector $\underline{x}$ are binary
$x_i \in \{+1, -1\}$. If the two networks produce an identical output
bit $\sigma^{A} = \sigma^B$, then their weights move one step in the
direction of $-x_i \sigma^A$. But the weights should remain in the
interval (\ref{acht}), therefore if any component moves out of this
interval, $|w_i| = L+1$, it is set back to the boundary $w_i = \pm L$.

Each component of the weight vectors performs a kind of random walk with
reflecting boundary. Two corresponding components $w_i^{A}$ and $w_i^B$
receive the same random number $\pm 1$. After each hit at the boundary
the distance $|w_i^{A} - w_i^B|$ is reduced until it has reached zero.
For two perceptrons with a $N$-dimensional weight space we have two
ensembles of $N$ random walks on the internal $\{ -L, ..., L\}$. If we
neglect the global signal $\sigma^{A} = \sigma^B$ as well as the bias
$\sigma^A$, we expect that after some characteristic time scale $\tau =
\mathcal{O}(L^2)$ the probability of two random walks being in different states
decreases as

\begin{equation}\label{zehn}
P(t) \sim P(0) e^{-t/\tau}
\end{equation}
Hence the total synchronization time should be given by $N \cdot P(t)
\simeq 1$ which gives

\begin{equation}\label{elf}
t_{\mathrm{sync}} \sim \tau \ln N
\end{equation}
In fact, our simulations for $N = 100$ show that two perceptrons with
$L=3$ synchronize in about 100 time steps and the synchronization time
increases logarithmically with $N$. However, our simulations also showed
that an opponent, recording the sequence of $(\sigma^{A} , \sigma^B,
\underline{x})_t$ is able to synchronize, too. Therefore, a single
perceptron does not allow a generation of a secret key.

\section{Secret key generation}
Obviously, a single perceptron transmits too much information. An
opponent, who knows the set of input/output pairs,  can derive the weights of
the two partners after synchronization. Therefore, one has to hide so
much information, that the opponent cannot calculate the weights, but on
the other side one has to transmit enough information that the two
partners can synchronize.

\begin{figure}
\sidecaption
\includegraphics[width=.6\textwidth]{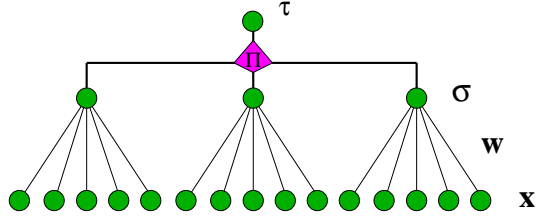}
\caption[]{Parity machine with three hidden units.}
\label{par}
\end{figure}
In fact, we found that multilayer networks with hidden units may be
candidates for such a task \cite{KaKiKa}. More precisely, we consider parity
machines with three hidden units as shown in Fig. \ref{par}. Each hidden unit is
a perceptron (\ref{eins}) with discrete weights (\ref{acht}). The
output bit $\tau$ of the total network is the product of the three bits
of the hidden units

\begin{eqnarray}\label{zwoelf}
\tau^{A} & = & \sigma^{A}_{1} \; \sigma^{A}_{2} \; \sigma^{A}_{3} \nonumber \\
\tau^B & = & \sigma^{B}_{1} \; \sigma^{B}_{2} \; \sigma^{B}_{3}
\end{eqnarray}
At each training step the two machines $A$ and $B$ receive identical
input vectors $\underline{x}_1 , \underline{x}_2 , \underline{x}_3$. The
training algorithm is the following: Only if the two output bits are
identical, $\tau^{A} = \tau^B$,  the weights can be changed. In this
case, only the hidden unit $\sigma_{i}$ which is identical to $\tau$
changes its weights using the Hebbian rule

\begin{equation}\label{dreizehn}
\underline{w}^{A}_{i} (t+1) = \underline{w}^{A}_{i} (t) - \underline{x}_i \tau^{A}
\end{equation}
For example, if $\tau^{A} = \tau^B = 1$ there are four possible
configurations of the hidden units in each network: 

\centerline{ $(+1, +1, +1), (+1,-1, -1), (-1, +1, +1), (-1, -1, +1)$}
\noindent
In the first case, all three weight vectors $\underline{w}_i,
\underline{w}_2, \underline{w}_3$ are changed, in all other three
cases only one weight vector is changed. The partner as well as any
opponent does not know which one of the weight vectors is updated.

The partners $A$ and $B$ react to their mutual stop and move signals
$\tau^{A}$ and $\tau^B$, whereas an opponent can only receive these
signals but not influence the partners with its own output bit.  This
is the essential mechanism which allows synchronization but prohibits
learning.  Numerical \cite{KaKiKa} as well as analytical \cite{RoKaKi}
calculations of the dynamic process show that the partners can
synchronize in a short time whereas an opponent needs a much longer
time to lock into the partners.

This observation holds for an observer who uses the same algorithm
(\ref{dreizehn}) as the two partners $A$ and $B$. Note that the observer
knows 1. the algorithm of $A$ and $B$,\, 2. the input vectors
$\underline{x}_1, \underline{x}_2, \underline{x}_3$ at each time step and \,
3. the output bits $\tau^{A}$ and $\tau^B$ at each time step.
Nevertheless, he does not succeed in synchronizing with $A$ and $B$
within the communication period.

\begin{figure}
\sidecaption
\includegraphics[width=.6\textwidth]{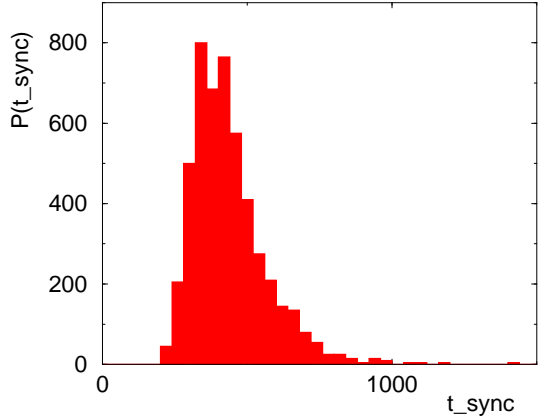}
\caption[]{Distribution of synchronization time for $N=100, L=3$.}
\label{tsync}
\end{figure}

\begin{figure}
\sidecaption
\includegraphics[width=.6\textwidth]{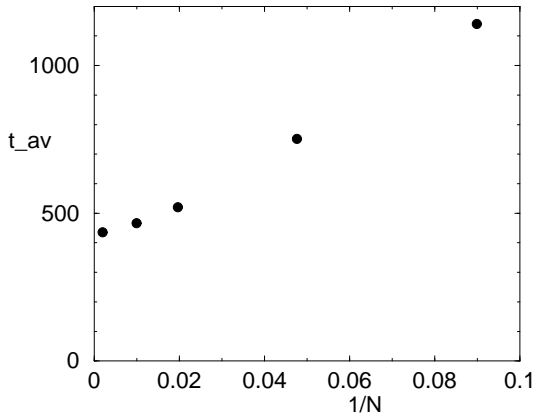}
\caption[]{Average synchronization time as a function of inverse system size.}
\label{tn}
\end{figure}
Since for each run the two partners draw  random initial weights and
since the input vectors are random, one obtains a distribution of
synchronization times as shown in Fig. \ref{tsync} for $N=100$ and $L=3$.
The average value of this distribution is shown  as a function of system
size $N$ in Fig. \ref{tn}. Even an infinitely large network needs only a
finite number of exchanged bits - about 400 in this case - to
synchronize, in agreement with the analytical calculation for $N
\rightarrow \infty$. 

\begin{figure}
\sidecaption
\includegraphics[width=.6\textwidth]{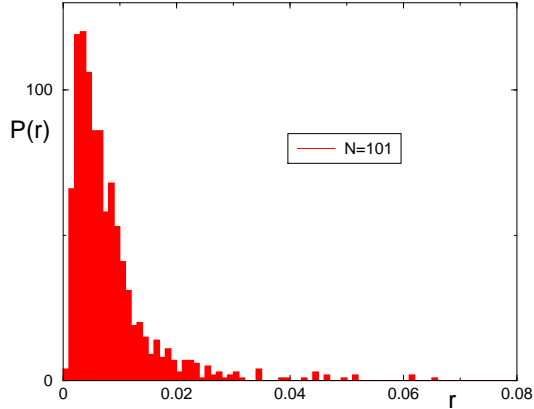}
\caption[]{
  Distribution of the ratio of synchronization time between networks A
  and B to the learning time of an attacker E.}
\label{rat}
\end{figure}

If the communication continues after
synchronization, an opponent has a chance to lock into the moving
weights of $A$ and $B$. Fig. \ref{rat} shows the distribution of the ratio
between the synchronization time of $A$ and $B$ and the learning time of the
opponent. In our simulations, for $N = 100$, this ratio never exceeded
the value $r=0.1$, and the average learning time is about 50000 time steps,
much larger than the synchronization time. Hence, the two partners can
take their weights $\underline{w}^{A}_{i} (t) = \underline{w}^{B}_{i}
(t)$ at a time step $t$ where synchronization most probably occurred as
a common secret key. Synchronization of neural networks can be used as a
key exchange protocol over a public channel.

\section{Conclusions}
Interacting neural networks have been calculated analytically. At each
training step two networks receive a common random input vector and
learn their mutual output bits. A new phenomenon has been observed:
Synchronization by mutual learning. If the learning rate $\eta$ is large
enough, and if the weight vectors keep normalized, then the two networks
relax to an antiparallel orientation. Their weight vectors still move
like a random walk on a hypersphere, but each network has complete
knowledge about its partner.

It has been shown how this phenomenon can be used for cryptography. The
two partners can agree on a common secret key over a public channel. An
opponent who is recording the public exchange of training examples
cannot obtain full information about the secrete key used for
encryption.

This works if the two partners use multilayer networks, parity machines.
The opponent has all the informations (except the initial weight vectors) of
the two partners and uses the same algorithms. Nevertheless he does not
synchronize.

This phenomenon may be used as a key exchange protocol. The two
partners select secret initial weight vectors, agree on a public
sequence of input vectors and exchange public bits. After a few steps
they have identical weight vectors which are used for a secret
encryption key. For each communication they agree on a new secret key,
without having stored any secret information before. In contrast to
number theoretical methods the networks are very fast; essentially they
are linear filters, the complexity to generate a key of length $N$
scales with $N$ (for sequential update of the weights).

Of course, one cannot rule out that algorithms for the opponent may be
constructed which find the key in much shorter time. In fact,
ensembles of opponents have a better chance to synchronize. In
addition, one can show that, given the information of the opponent,
the key is uniquely determined, and, given the sequence of inputs, the
number of keys is huge but finite, even in the limit $N \rightarrow
\infty$ \cite{Ur}.  These may be good news for a possible attacker.
However, recently we have found advanced algorithms for
synchronization, too. Such variations are subjects of active research,
and future will show whether the security of neural network
cryptography can compete with number theoretical methods.

{\bf Acknowledgments}: This work profitted from enjoyable
collaborations with Richard Metzler and  Michal Rosen-Zvi.
We thank the German Israel Science Foundation (GIF) and the Minerva
Center of the Bar-Ilan University for support.

\end{document}